# First-principles thermal equation of state and thermoelasticity for hcp Fe under high pressures

Xianwei Sha and R. E. Cohen

Carnegie Institution of Washington, 5251 Broad Branch Road, NW, Washington, D. C. 20015, U. S. A.

We investigate the equation of state and elastic properties of nonmagnetic hcp iron at high pressures and high temperatures using the first principles linear response linear-muffin-tin-orbital method in the generalized-gradient approximation. We calculate the Helmholtz free energy as a function of volume, temperature, and volume-constrained strain, including the electronic excitation contributions from band structures and lattice vibrational contributions from quasi-harmonic lattice dynamics. We perform detailed investigations on the behavior of elastic moduli and equation of state properties as a function of temperature and pressure, including the pressure-volume equation of state, bulk modulus, the thermal expansion coefficient, the Grüneisen ratio, and the shock Hugoniots. A detailed comparison has been made with available experimental measurements and theoretical predictions.

## I. Introduction

Iron is one of the most abundant elements in the Earth, and is fundamental to our world. The study of iron at high pressures and temperatures is of great geophysical interest, since both the Earth's liquid outer core and solid inner core are mainly composed of this element. Although the crystal structure of iron at the extremely high temperature (4000 to 8000 K) and high pressure (330 to 360 GPa) conditions found in the deep inner core is still under debate,[1-5] the hexagonal-close-packed phase ($\varepsilon$-Fe) has been found to have a wide stability field extending from deep mantle to core conditions, and serves as a starting point for understanding the nature of the inner core.[6] Significant experimental and theoretical efforts have been recently devoted to investigate properties of hcp iron at high pressures and temperatures. New high-pressure diamond-anvil-cell techniques have been developed or significantly improved, which makes it able to reach higher pressures and provide more information on material properties in these extreme states. First-principles based theoretical techniques have been improved in reliability and accuracy, and have been widely used to predicate the high pressure-temperature behavior and provide fundamental understandings to the experiment.

Despite the intensive investigations, numerous fundamental problems remain unresolved, and many of the current results are mutually inconsistent.[6] The melting line at very high pressures has been one of the most difficult and controversial problems.[7-14] Other major problems include possible subsolidus phase transitions[2, 4-6, 15] and the magnetic structure of the dense hexagonal iron.[16-18] There are also many debates about the lattice strains at high temperatures. Two earlier calculations used the particle-in-cell (PIC) model to obtain the lattice vibrational contributions, and both predicted a rapid increase in the *c/a* axial ratio to above 1.7 at the core conditions.[19, 20] However, later theoretical work[21-24] and experiment[7] gave much smaller temperature dependences. We recently reported that the results from the first-principles linear response calculations and the PIC model usually agree well except when the lattice approaches instability, and both theoretical techniques predicted a slight increase in the axial ratio with temperature.[25] Since the on-site anharmonicity is small up to the melting temperature,[25] we further use the first-principles linear response calculations based on the full-potential linear-muffin-tin-orbital (LTMO) method and quasi-harmonic approximation to examine the thermal equation of state of nonmagnetic hcp Fe.

Knowledge about the elasticity of hcp iron and its pressure and temperature dependences plays a crucial role in understanding the seismological observations of the inner core, such as the low shear velocity and the elastic anisotropy. Several sets of first-principles elastic moduli have been reported for hcp Fe at high pressures[20, 26-29], most at zero temperature. Steinle-Neumann et al. examined the thermoelasticity at the inner core condition using first-principles pseudopotential calculations and the PIC model.[20] However, their calculations gave too large *c/a* ratios at high temperatures. Most recently Vocadlo examined the elastic constants of iron and several iron alloys at high temperatures via *ab initio* molecular dynamics simulations and thermodynamic integration, but she only reported the results for hcp Fe at two high temperatures and atomic densities due to the intensive computational requirements for such calculations.[30] Here we present the calculated elasticity of nonmagnetic hcp Fe as a function of pressure and temperature using first-principles linear response calculations.

In section II we detail the theoretical methods to perform the first-principles calculations and obtain the thermal properties and elastic moduli. We present the results and related discussions about the thermal equation of state in section III, and about the thermoelasticity in section IV. We conclude with a brief summary in Section V.

## II. Theoretical methods

The Helmholtz free energy F for many metals has three major contributions[31]

$$F(V,T,\delta)=E_{static}(V,\delta)+F_{el}(V,T,\delta)+F_{ph}(V,T,\delta) \qquad (1)$$

with V as the volume, T as the temperature, and δ as the strain. $E_{static}$ is the zero-temperature energy of a static lattice, $F_{el}$ is the thermal free energy arising from electronic excitations, and $F_{ph}$ is the lattice vibrational energy contribution. We obtain both $E_{static}$ and $F_{el}$ from the first-principles calculations directly, assuming that the eigenvalues for given lattice and nuclear positions are temperature-independent and only the occupation numbers change with temperature through the Fermi-Dirac distribution.[19, 25, 32] The calculated electronic entropies of nonmagnetic hcp Fe using the full-potential LMTO method and static eigenvalues agree very well with the values from self-consistent high temperature Linear-Augmented-Plane-Wave (LAPW) method[19] over a wide temperature (6000-9000K) and volume (40-90 bohr$^3$/atom) range. The linear response method gives the phonon dispersion spectrum and phonon density of states, which provide both a microscopic basic for and a means of calculating the thermodynamic and elastic properties.[6] We obtain the vibrational free energy within the quasiharmonic approximation. Earlier theoretical calculations using thermodynamic integration and the PIC model both indicate that the on-site anharmonicity in hcp Fe is small up to the melting temperatures.[21, 25]

The computational approach is based on the density functional theory and density functional perturbation theory, using multi-κ basis sets in the full-potential LMTO method.[33, 34] The induced charge densities, the screened potentials and the envelope functions are represented by spherical harmonics inside the non-overlapping muffin-tin spheres surrounding each atom and by plane waves in the remaining interstitial region. We use the Perdew-Burke-Ernzerhof (PBE) generalized-gradient-approximation (GGA) for the exchange and correlation energy.[35] The **k**-space integration is performed over a 12×12×12 grid using the improved tetrahedron method.[36] We use the perturbative approach to calculate the self-consistent change in the potential,[37, 38] and determine the dynamical matrix for a set of irreducible **q** points on a 6×6×6 reciprocal lattice grid. Careful convergence tests have been made against **k** and **q** point grids and other parameters. We examine hcp Fe at volumes from 40 to 80 bohr$^3$/atom and at c/a ratios from 1.5 to 1.7 in 0.05 interval. We determine the equilibrium thermal properties by minimizing the Helmholtz free energies with *c/a* ratio at a given temperature and volume.

We obtain the elastic moduli as the second derivatives of the Helmholtz free energies with respect to strain tensor, by applying volume-conserving strains and relaxing the symmetry-allowed internal coordinates. For hexagonal crystals, the bulk modulus K and shear modulus $C_S$ yield the combinations of the elastic moduli

$$K = [C_{33}(C_{11} + C_{12}) - 2C_{13}^2]/C_S \qquad (2)$$

$$C_S = C_{11} + C_{12} + 2C_{33} - 4C_{13} \qquad (3)$$

To make direct comparisons to the ultrasonic measurements, we use the adiabatic bulk modulus $K_s$[31]

$$K_S = (1 + \alpha\gamma T) \times K_T \qquad (4)$$

where $K_T$ is the isothermal bulk modulus, α is the thermal expansivity, and γ is the Grüneisen parameter. We obtain the equation of state parameters $K_T$, α, and γ as functions of temperature and pressure from the first-principles linear response calculations.

We calculate $C_S$ by varying the *c/a* ratio at a given volume:

$$\varepsilon(\delta) = \begin{pmatrix} \delta & 0 & 0 \\ 0 & \delta & 0 \\ 0 & 0 & (1+\delta)^{-2} - 1 \end{pmatrix} \tag{5}$$

where δ is the strain magnitude. The Helmholtz free energy F(δ) is related to δ as:

$$F(\delta) = F(0) + C_S V \delta^2 + O(\delta^3) \tag{6}$$

with F(0) as the free energy of the unstrained structure.

The volume dependences of the equilibrium *c/a* ratio are related to the difference in the linear compressibility along the *a* and *c* axes

$$-\frac{d \ln(c/a)}{d \ln V} = (C_{33} - C_{11} - C_{12} + C_{13})/C_S \tag{7}$$

We apply a volume-constrained orthorhombic strain to calculate the difference between $C_{11}$ and $C_{12}$, $C_{11}-C_{12}=2C_{66}$,

$$\varepsilon(\delta) = \begin{pmatrix} \delta & 0 & 0 \\ 0 & -\delta & 0 \\ 0 & 0 & \delta^2/(1-\delta^2) \end{pmatrix} \tag{8}$$

The corresponding free energy change is:

$$F(\delta) = F(0) + 2C_{66} V \delta^2 + O(\delta^4) \tag{9}$$

We use a monoclinic strain to determine $C_{44}$

$$\varepsilon(\delta) = \begin{pmatrix} 0 & 0 & \delta \\ 0 & \delta^2/(1-\delta^2) & 0 \\ \delta & 0 & 0 \end{pmatrix} \quad (10)$$

which leads to the energy change

$$F(\delta) = F(0) + 2C_{44}V\delta^2 + O(\delta^4) \quad (11)$$

When evaluating $C_{44}$ and $C_{66}$, we relax the internal degree of freedom by minimizing the total energy with respect to the atomic positions in the two atom primitive unit cell.[28, 39] Since the leading error term is third order in δ for $C_s$ and fourth order for $C_{44}$ and $C_{66}$, we include both positive and negative strains to calculate $C_s$. We choose 4-6 values for each strain ranging from 0 to 0.03, and perform first-principles linear response calculations to obtain the band structure and phonon density of states for all the strained structures at each volume. We then calculate the Helmholtz free energies at temperatures from 0 to 6000 K, and fit a polynomial of the free energies to the strain magnitudes. The quadratic coefficients of the polynomial fitting give the elastic moduli that appear in the equations of motion and directly give sound velocities.[39-41]

### III. Thermal equation of state

We present in Fig. 1 the calculated phonon density of states (DOS) of hcp Fe at the c/a ratio of 1.6 and volumes of 40, 60, and 70 bohr$^3$/atom. Nuclear resonant inelastic x-ray scattering techniques have been used to examine the phonon DOS up to high pressures,[42-45] and our first-principles linear response results show good agreements with these measurements. The Raman-active $E_{2g}$ phonon correlates with the zone–edge acoustic mode, the elastic modulus $C_{44}$, and shear-wave velocity, and their frequencies at high pressures have been recently measured using the Raman spectroscopy.[46, 47] Our linear-response $E_{2g}$ frequencies show excellent agreements with the first-principles frozen-phonon values[16, 46] at both ambient and high pressures, as shown in Fig. 2. Although the theory gave similar pressure dependences of the Raman frequencies as experiment,[46, 47] all these theoretical calculations overestimate the $E_{2g}$ frequencies. At low pressures, the antiferromagnetic nature of the ground state hcp Fe leads to splitting of the Raman frequencies,[16] as well as substantial temperature and compositional dependence.[47] All these account for the discrepancies between the theory and experiment.

We fit the calculated Helmholtz free energies for each given temperature to some equation of state (EoS) formulations to obtain the bulk modulus and thermal pressures. Due to its versatility and high accuracy, we choose the Vinet EoS form[48-50]

$$F(V,T) = F_0(T) + \frac{9K_0(T)V_0(T)}{\xi^2}\{1 + \{\xi(1-x) - 1\}\exp\{\xi(1-x)\}\}] \quad (12)$$

where $F_0(T)$ and $V_0(T)$ are the zero pressure equilibrium energy and volume, $x = (V/V_0)^{1/3}$, $K_0(T)$ is the bulk modulus, $\xi = \frac{3}{2}(K_0' - 1)$ and $K_0' = [\frac{\partial K(T)}{\partial P}]_0$. The subscript 0 throughout represents the standard state P= 0 GPa. We list the calculated Vinet EoS parameters at ambient condition in Table I. The current LMTO results agree well with recent first principles calculations for nonmagnetic hcp Fe using the all-electron Linearized-Augmented-Plane-Wave (LAPW) Method[39] and the Projector-Augmented-Wave (PAW) method,[51] both using the PBE GGA functional. The discrepancy between the nonmagnetic calculations and diamond-anvil-cell experiments[52-54] is significantly larger for hcp Fe than for typical transition metals. As shown in earlier calculations, at low pressures (P < 50 GPa) hcp Fe might have antiferromagnetic structure, and including the magnetism significantly improves the agreements with experiment.[39]

The temperature dependences of the Vinet EoS parameters $V_0(T)$, $K_0(T)$, and $K_0'(T)$ are plotted in Fig. 3, which gave similar trends as in bcc Fe[32] and Ta.[55]

We obtain the pressure analytically from the Vinet EoS parameters:

$$P(V,T) = \{\frac{3K_0(T)(1-x)}{x^2}\}\exp\{\xi(1-x)\} \qquad (13)$$

In Fig. 4 we show the calculated pressure-volume equation of state for hcp Fe at temperatures between 0 to 3000 K in 500 K intervals. Compared to the ambient-temperature x-ray diffraction measurements, our first-principles results agree well with the experiments to 78 GPa with Ar and Ne pressure-transmitting media[56] and to 304 GPa without a medium.[52] The discrepancies between our calculated data and the experiment are larger at low pressures, consistent with earlier calculations.[39] Spin-polarized GGA calculations for an antiferromagnetic structure agree better with the experiment at low pressures.[39]

We obtain the thermal pressures as a function of volume and temperature according to the pressure differences among the EoS isotherms, as shown in Fig. 5. The thermal pressures are small and volume-independent at low temperatures, and increase significantly and show complex volume-dependence with further increase in temperature. At a given volume, the thermal pressures increase linearly with temperature. All these are similar to the behavior in bcc Fe[32] and bcc Ta.[55]

In the Debye approximation, an accurate high-temperature global equation of state can be formed from the 0 K Vinet isotherm plus a volume-dependent thermal free energy $F_{th}$ [31],

$$F_{th} = RT[\frac{9}{8}(\frac{\theta_D}{T}) + 3\ln(1-e^{-\theta_D/T}) - D(\frac{\theta_D}{T})] \qquad (14)$$

Debye function $D(\theta_D/T)$ is

$$D(\frac{\theta_D}{T}) = 3(\frac{T}{\theta_D})^3 \int_0^{\theta_D/T} \frac{z^3 dz}{e^z - 1} \tag{15}$$

We calculate the Debye temperatures $\theta_D(T)$ at 0K by numerical integration of low-frequency part of the phonon density of state, and solved Eqn. 14 to obtain $\theta_D(T)$ at other temperatures. The calculated and fitted thermal free energies agree well at different temperatures and volumes with rms deviations of ~0.4 mRy.

We derive various thermal equation of state properties analytically from the Helmholtz free energy.[31] The thermal expansion coefficient α is:

$$\alpha = -\frac{1}{V}(\frac{\partial^2 F}{\partial T \partial V})/(\frac{\partial^2 F}{\partial V^2})_T = \frac{3R\gamma_D}{K_T V}[4D(\frac{\theta}{T}) - \frac{3(\theta/T)}{e^{\theta/T} - 1}] \tag{16}$$

The calculated α increases linearly with temperature at both ambient and high pressures (Fig. 6), as predicted by the quasiharmonic approximation at the high T limit.[31] The calculations show fairly good agreements with the shock wave[57] and in situ x-ray[53] measurements at high pressures and temperatures (P ≈ 200 GPa, T ≈ 5200 K). Isaak and Anderson determined the thermal expansivity of hcp Fe at high pressures and temperatures based on the reported experimental properties using thermodynamic approximations.[58] Compared to their thermodynamic analysis, our first-principles calculations give much better agreements with the shock and in situ x-ray data.

The Anderson-Grüneisen parameter $\delta_T$ is used to characterize the pressure dependence of the thermal expansion coefficient:

$$\delta_T = (\frac{\partial \ln \alpha}{\partial \ln V})_T = -\frac{1}{\alpha K_T}(\frac{\partial K_T}{\partial T})_P \tag{16}$$

The calculated $\delta_T$ shows complex temperature dependences at different pressures, and drops rapidly with increasing pressure at a given temperature, as shown in Fig. 7(a). For many materials, the parameter $\delta_T$ can be fitted as a function of volume:

$$\delta_T = \delta_T(\eta = 1) \times \eta^\kappa$$

(17)

where η=V/V$_0$(T$_0$). The equation works well for transition metals such as bcc Ta[55] and metal oxides such as MgO.[59] As shown in Fig. 7(b), although δ$_T$ of hcp Fe shows a strong decrease with compression, it does not drop as rapidly as power order at high pressures, similar to what has been observed in bcc Fe.[32]

The Grüneisen ratio γ is an important parameter in understanding the relationship between the thermal and elastic properties:

$$\gamma = V\left(\frac{\partial P}{\partial U}\right)_V = \frac{\alpha K_T V}{C_V} = V\frac{\partial^2 F}{\partial V \partial T} \bigg/ \left(\frac{\partial U}{\partial T}\right)_V \quad (18)$$

where U is the internal energy. Many different techniques have been used to determine the Grüneisen ratio of hcp Fe, including the nuclear resonant inelastic x-ray scattering,[43] Raman,[46] x-ray diffraction,[53, 60] shock wave,[61] and thermodynamic analysis.[62, 63] At a given pressure, the first-principles calculated γ first increases with temperature, and then drops rapidly at high temperatures (T >1500 K), as shown in Fig. 8. The pressure dependence of γ is complex and strongly temperature dependent. The calculated ratios at 500K agree fairly with ambient-temperature x-ray diffraction measurements. The volume dependence of the Grüneisen ratio is defined by the parameter *q*:

$$q = \frac{\partial \ln \gamma}{\partial \ln V} \quad (19)$$

The parameter *q* is usually treated as a constant, and its experimental value for hcp Fe varies from 0.6 to over 1.6 depending on the pressure range and measuring methods. Our first-principles calculations show that *q* strongly depends on both the temperature and pressure, and can even be negative at some pressure and temperature regimes [Fig. 9]. Similar complex behavior has been reported for bcc Ta[55] and bcc Fe[32] before.

Shock compression experiment is usually used to obtain the high-pressure high-temperature equation of state along the Hugoniot. We calculate the relationship between the pressures $P_H$ and temperatures $T_H$ along the Hugoniot according to the Rankine-Hugoniot equation:

$$\frac{1}{2}P_H[V_0(T_0) - V] = E_H - E_0(T=0) \tag{20}$$

E is the internal energy. We obtain the $P_H$ and $T_H$ based on our thermal equation of state results by varying the temperature at a given volume until the Rankine-Hugoniot equation is satisfied. The calculated data agree with experiment for both the shock Hugoniot[64] and the temperatures along the Hugoniot,[65], as well as earlier estimations using plausible bounds for specific heat and experimental constraints for the Grüneisen parameter,[66] as shown in Fig. 10.

IV.     **Thermoelasticity**

Many different sets of experimental[67-75] and theoretical[20, 26, 28, 29, 39, 51, 76-80] elastic moduli have already been reported for ε-Fe. We present our calculated static moduli of nonmagnetic hcp Fe as a function of volume in Fig. 11, in comparison to several pervious experimental and theoretical data. One of the major reasons for the wide distribution of the experimental data is that single-crystal elastic moduli extracted from radial x-ray diffraction data on polycrystalline samples under nonhydrostatic compression contain large errors, since the assumption of a single uniform macroscopic stress applied to all grains is violated due to plastic deformation[81, 82]. As shown for another hcp transition metal cobalt, the $C_{11}$, $C_{33}$, $C_{12}$ and $C_{13}$ are 20% off with respect to single-crystal measurements, and the discrepancies are up to 50% and 300% for shear moduli $C_{66}$ and $C_{44}$ [81]. Our calculated elastic moduli show a strong increase with the increase of pressure, and show fair agreements with experiment and earlier theoretical calculations, except at low pressures. The large discrepancies between theory and experiment at low pressures might be attributed to the anti-ferromagnetic nature of hcp Fe, which is predicted to vanish at pressures higher than 50 GPa.[16]

Most of the previous experiment and theoretical calculations only focus on the elastic properties at zero or ambient temperatures, without thermal effects included. Only very recently, people began to examine the temperature effects on sound velocities using nuclear inelastic x-ray scattering in a laser-heated diamond anvil cell.[71] Steinel-Neumann et al. performed first principles calculations using a plane wave mixed basis method and PIC model to examine the elasticity at the Earth's inner core conditions.[20] However, their high-temperature results are questionable since their calculations gave too large increases in the c/a axial ratios compared to other theoretical and experimental predictions.[25] In Fig. 12 we show our calculated elastic moduli as a function of temperature at several different volumes, in comparison to the theoretical data of Steinle-Neumann et al.[20] At a given atomic volume, our calculated elastic moduli show modest change with the temperature in a quite linear manner. Most of the moduli show different temperature dependences due to the large differences in the c/a ratios at high temperatures. Most recently Vocadlo reported that the elastic moduli of bcc Fe and iron alloys do not show any significant variation with temperature at a given atomic density,[30] similar to what we have observed for hcp Fe here. We interpolate our high-pressure high-temperature moduli to obtain the elastic properties at the two temperatures that Vocadlo examined for ε-Fe. $C_{13}$ and $C_{33}$ agree in ~5%, $C_{11}$ and $C_{12}$ usually agree within ~10%. However, the differences between the predicted shear moduli $C_{66}$ and $C_{44}$ are pretty large, up to 15% and 35%. Our zero-temperature shear moduli agree well with Vocadlo's earlier work[29], so the differences come from the thermal parts. We use linear response lattice dynamics, and Vocadlo used *ab initio* molecular dynamics and thermodynamic integration to obtain the thermal contributations. One major difference is that our calculations are using quasi-harmonic approximations. As shown in earlier calculations using both thermodynamic integration and the PIC model[21, 25], the on-site anharmonicity in ε-Fe is small up to the melting temperature. The discrepancies might also come from the different set-ups in the first-principles calculations. Vocadlo used a 64-atom supercell and 4 irreducible *k* points in her *ab initio* molecular dynamics simulations. We carefully compare the calculated $C_{44}$ values at different *k* point meshes up to 24×24×24 and *q* meshes up to 6×6×6 , and make sure our results are converged. Raman spectroscopy at high temperatures and pressures might give critical information to to understand the temperature dependences of $C_{44}$ and resolve the discrepancy between the two calculations.

## V.   Conclusions

In summary, we present the thermal equation of state properties and thermoelasticity of nonmagnetic hcp Fe at high pressures from first-principles linear response calculations. The calculated lattice dynamics at high pressures agrees with nuclear resonant inelastic x-ray scattering and Raman measurements. The pressure-volume equation of state agrees well with experiment at high pressures. The thermal expansion coefficient at high pressures and temperatures shows much better agreement with the experiment than the values predicted from thermodynamic analysis. The calculated Grüneisen ratio and shock Hugoniot agree with the experiment. The parameter $q$, which is usually considered as a constant, shows strong temperature and pressure dependences. The calculated static elastic and bulk moduli at ambient temperature are in fairly good agreements with measurements and previous calculations. At a given atomic volume, the elastic moduli show modest changes with temperature in a quite linear manner.

## Acknowledgements

We thank Dr. S. Y. Savrasov for kind agreement to use his LMTO codes and many helpful discussions. This work was supported by DOE ASCI/ASAP subcontract B341492 to Caltech DOE w-7405-ENG-48. Computations were performed on the Opteron Cluster at the Geophysical Laboratory and ALC cluster at Lawrence Livermore National Lab, supported by DOE and the Carnegie Institution of Washington.

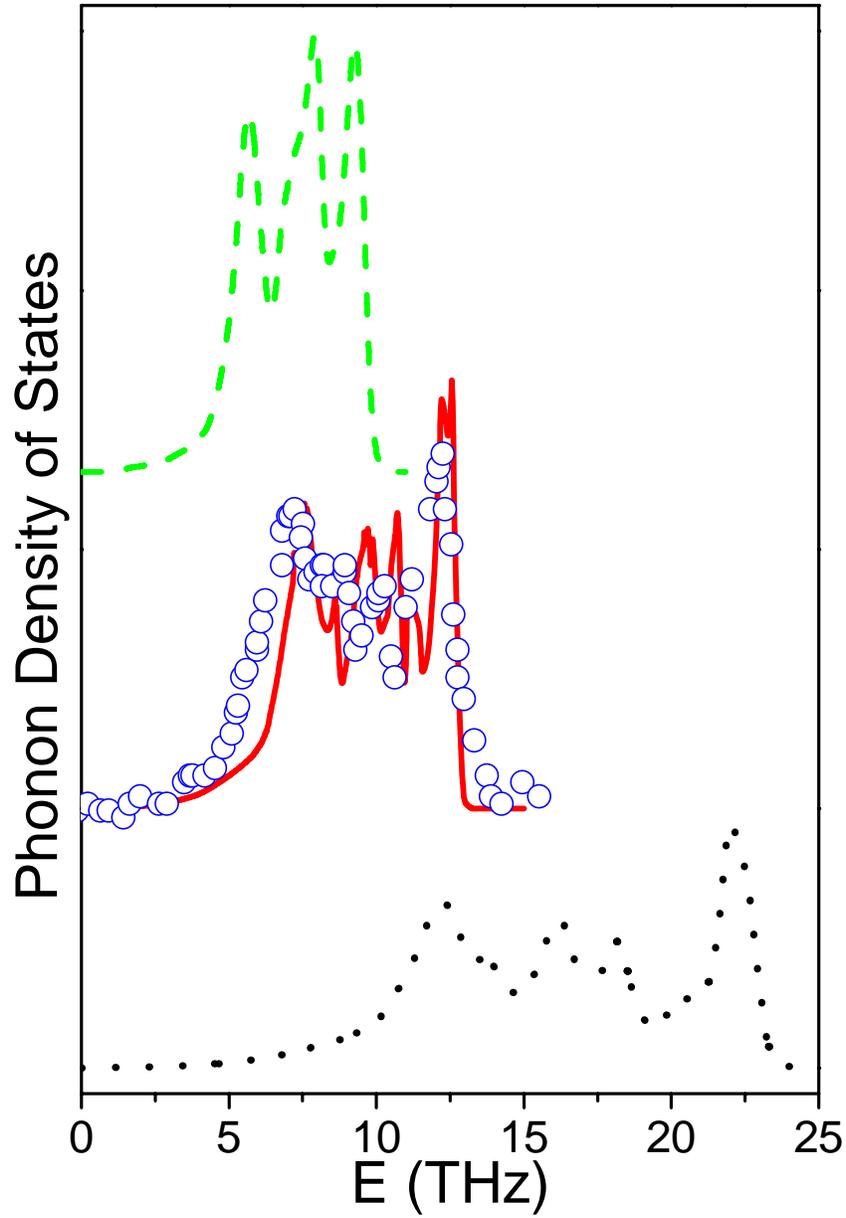

Fig. 1 The calculated phonon density of states for nonmagnetic hcp Fe at *c/a* ratio of 1.6 and volumes of 40, 60 and 70 bohr$^3$/atom, shown as the dotted, solid and dashed lines respectively. The computed data at 60 bohr$^3$/atom agree with the nuclear resonant inelastic x-ray scattering measurements at 50GPa (dots, Ref. 41), which gives a similar experimental volume according to the equation of state.

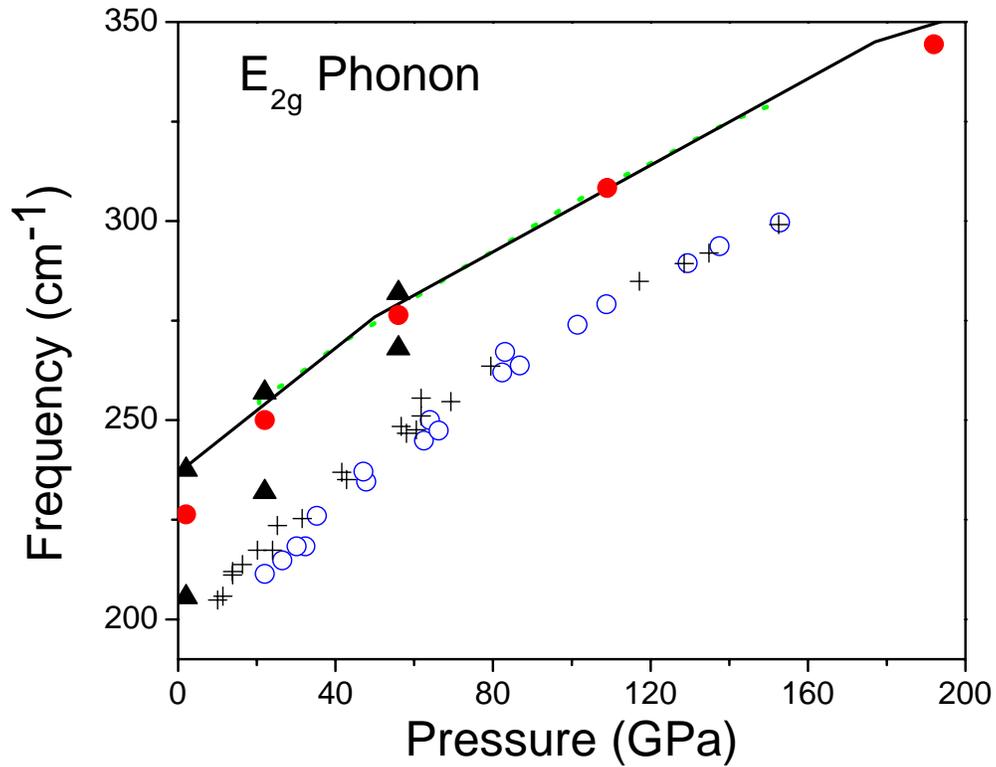

Fig. 2  The pressure dependence of the Raman phonon frequencies for hcp Fe.  The linear-response calculated frequencies (solid line) show excellent agreements with frozen-phonon calculations (dotted line, Ref. 45; filled circles, Ref. 16).  Results from earlier theoretical calculations for antiferromagnetic hcp Fe (filled triangles, Ref. 16) and two Raman measurements (open circles, Ref. 45; cross, Ref. 46) are also shown.

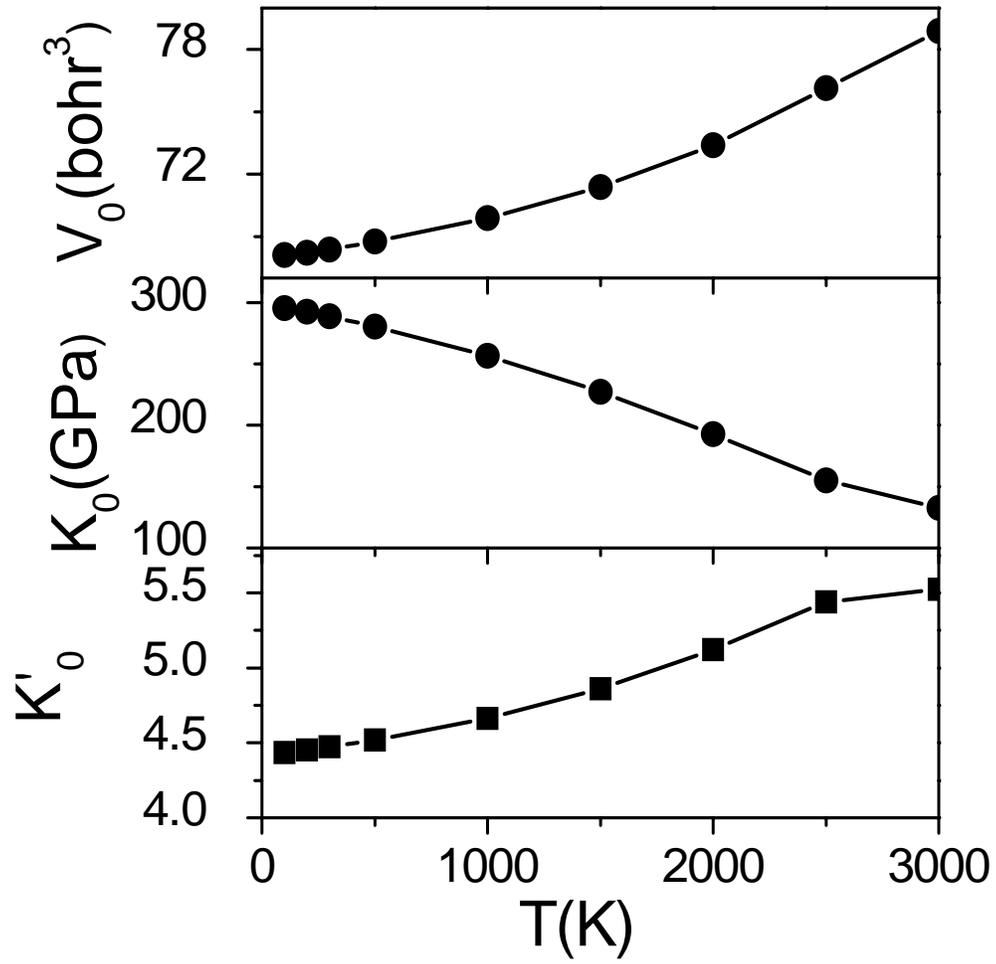

Fig. 3  Fitted Vinet equation of state parameters $V_0(T)$, $K_0(T)$, and $K_0'(T)$ as a function of temperature for hcp Fe.

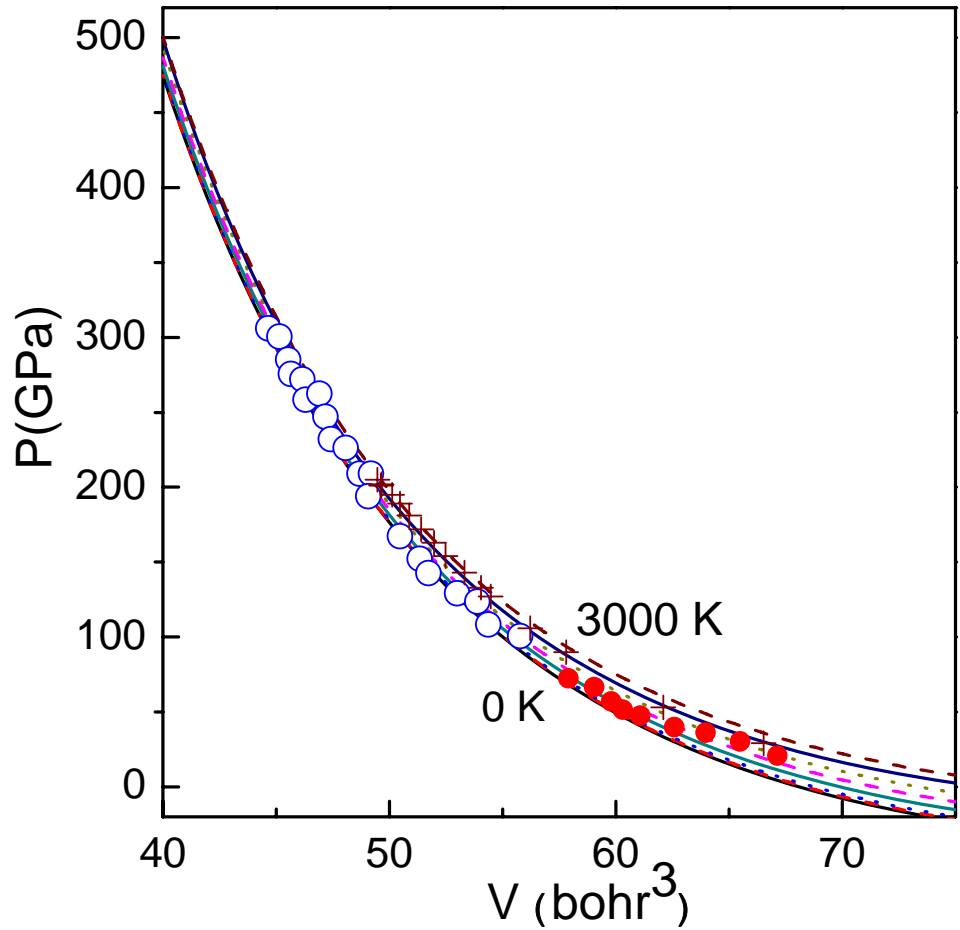

Fig. 4 The calculated pressure–volume equation of state (lines) for hcp Fe at temperatures from 0 to 3000 K in 500 K intervals. The ambient-temperature results agree with the diamond-anvil-cell x-ray diffraction measurements at high pressures (filled circles, Ref. 55; open circles, Ref. 52, cross, Ref. 54)

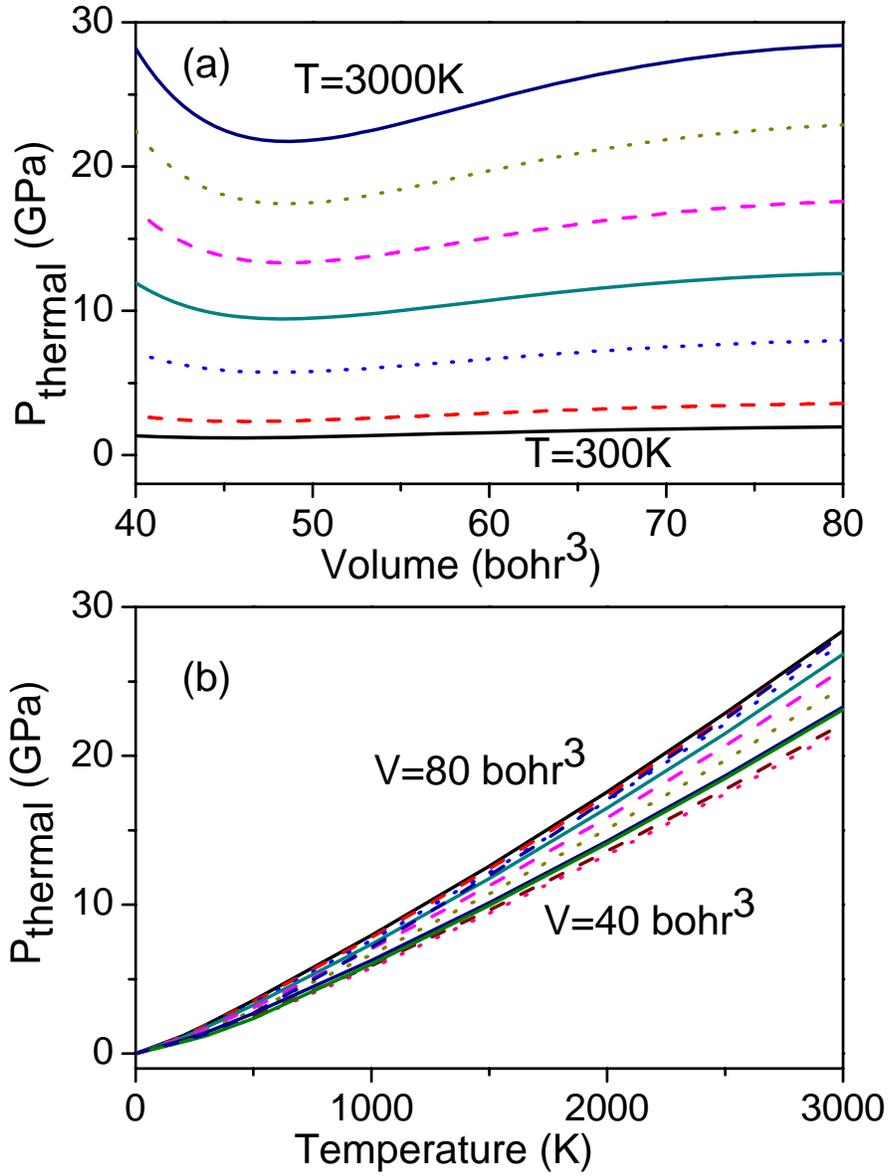

Fig. 5 The calculated thermal pressures for hcp Fe as a function of volume at temperatures of 300 K and from 500 K up to 3000 K in 500 K intervals (a), and as a function of temperature for volumes varying from 40 to 80 bohr$^3$/atom (b).

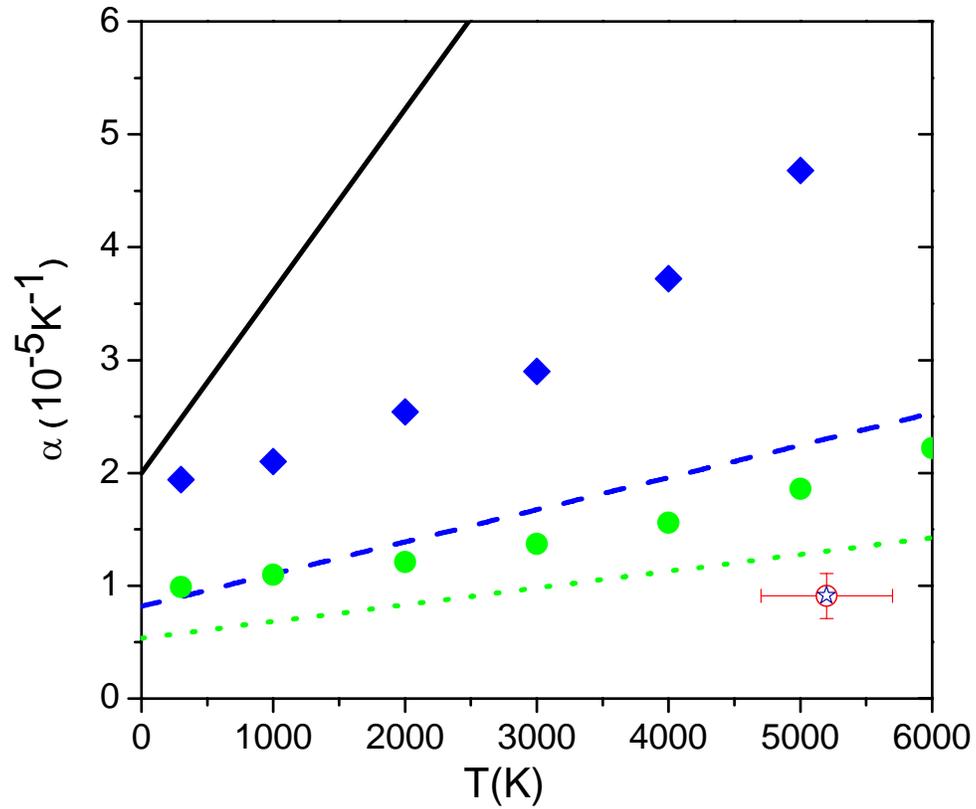

Fig. 6 The calculated thermal expansion coefficients of hcp Fe as a function of temperature at 0, 100 and 200 GPa, shown as the solid, dashed and dotted lines, in comparison to the results from shock compression at 202±3 GPa (open circle with error bar, Ref. 56), in situ x-ray measurement at 202 GPa (star, Ref. 53), and thermodynamic analysis at 100 (filled diamonds, Ref. 57) and 200 (filled circles, Ref. 57) GPa.

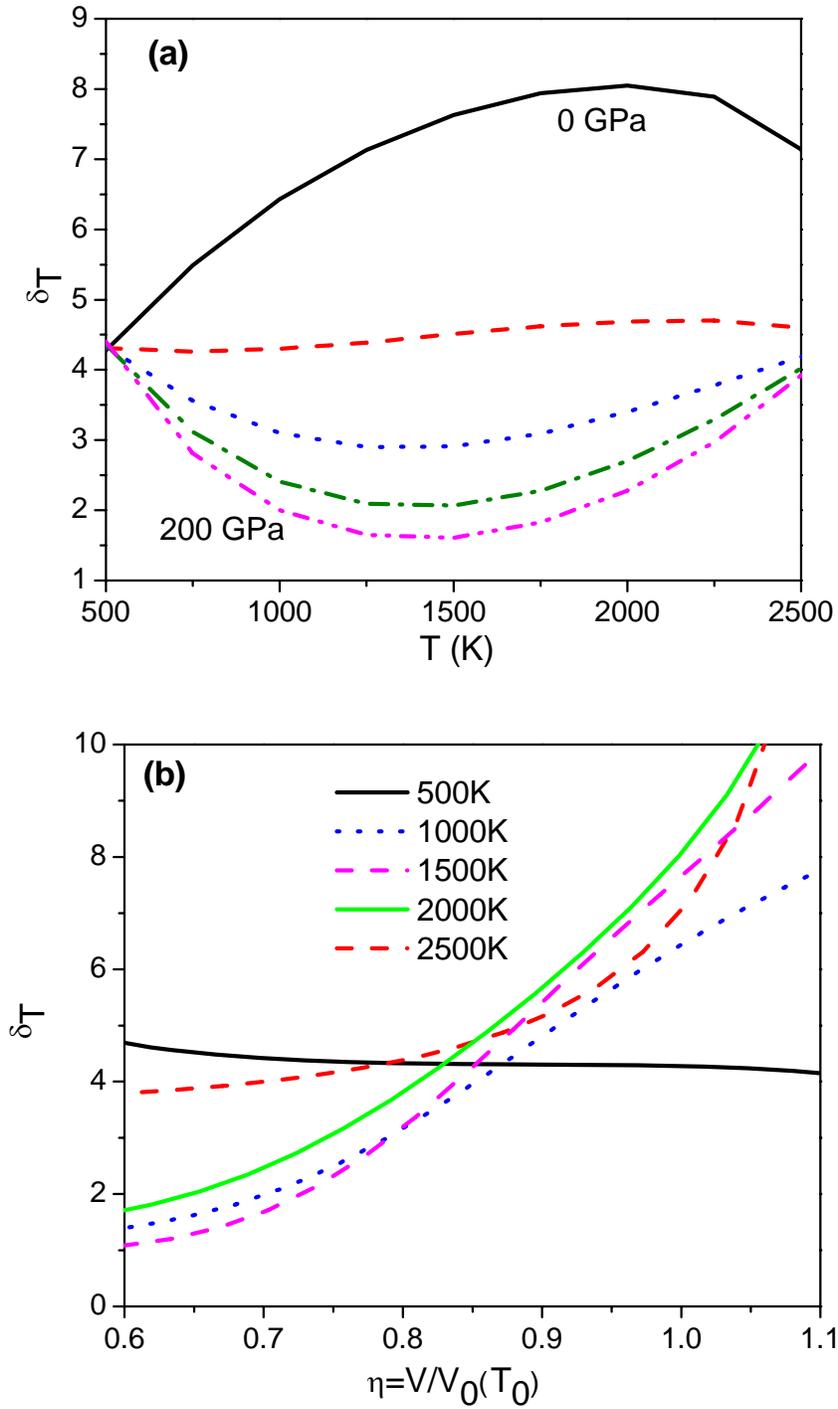

Fig. 7 The Anderson-Grüneisen parameter $\delta_T$ as a function of temperature at pressures from 0 to 200 GPa in 50 GPa intervals (a), and as a function of volume (b).

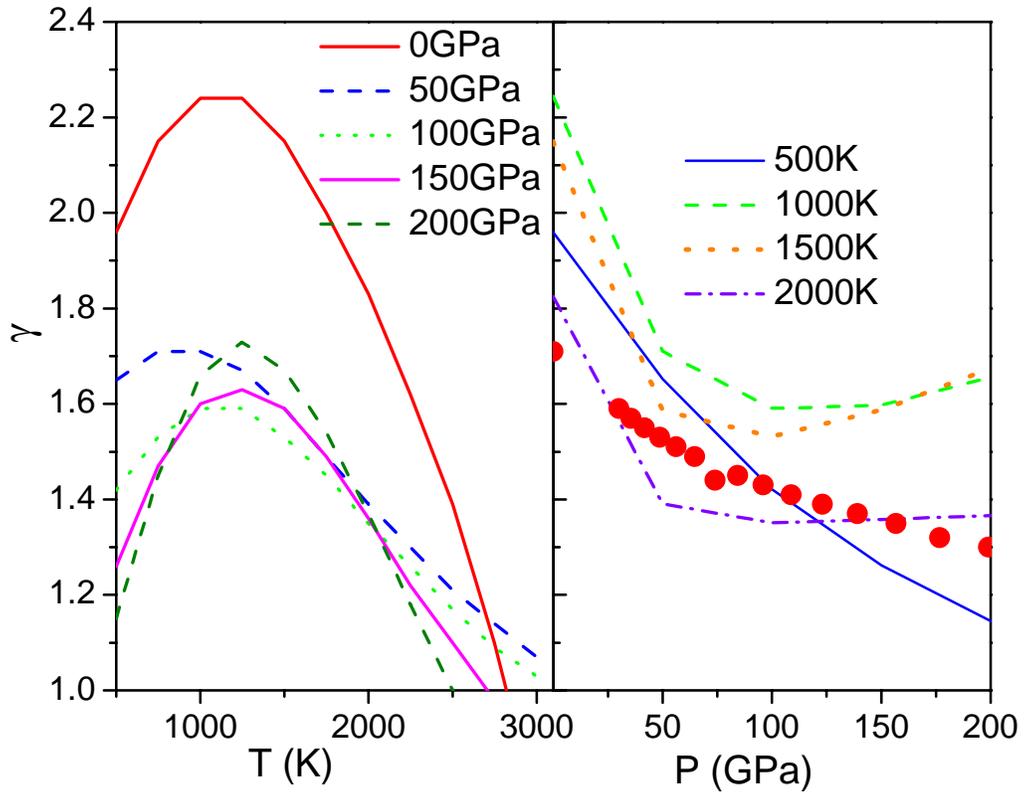

Fig. 8 The Grüneisen ratio γ of hcp Fe as a function of temperature (a) and pressure (b). Ambient-temperature x-ray diffraction data (filled circles) are from Ref. 53 and 59.

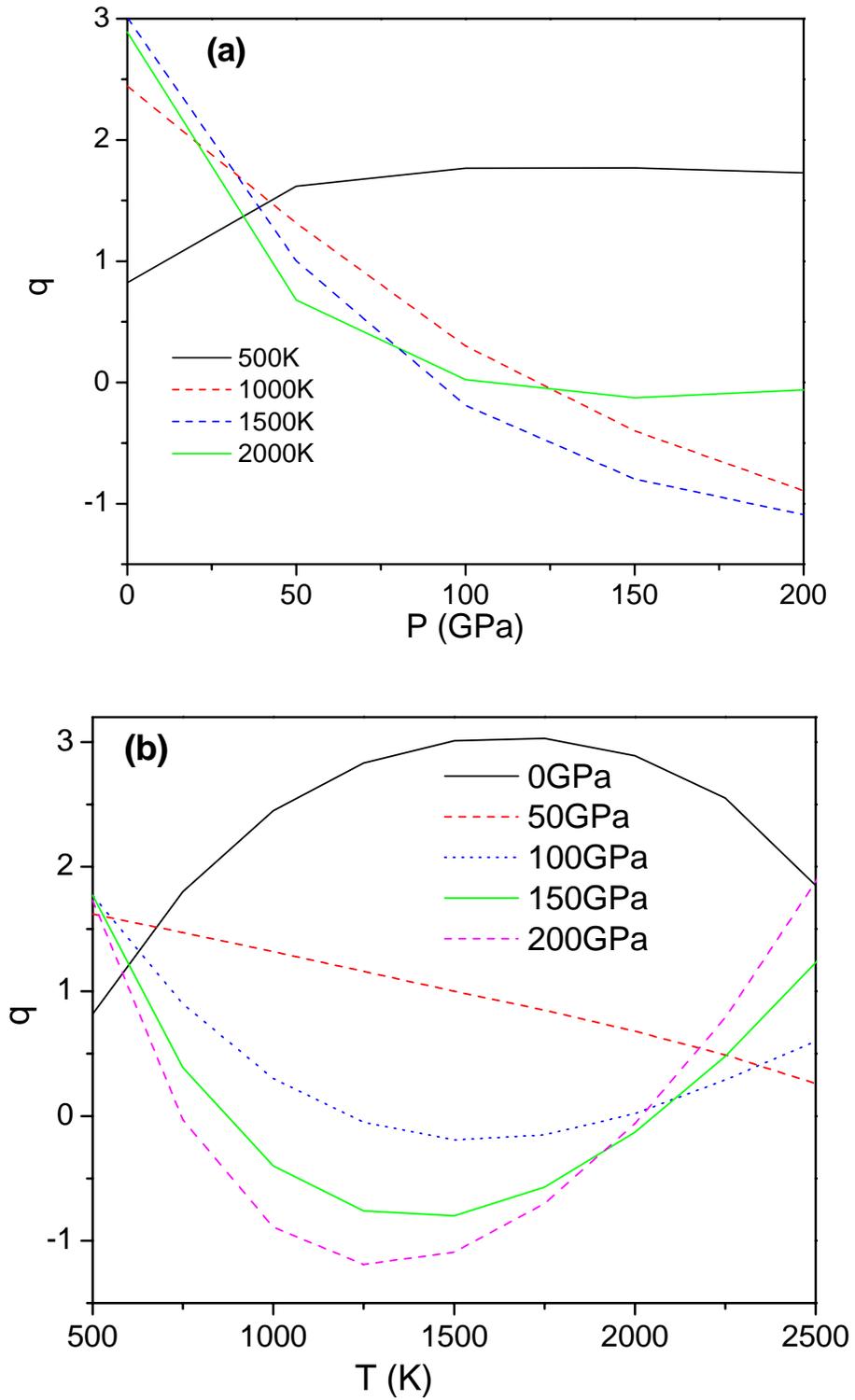

Fig. 9 The pressure (a) and temperature (b) dependences of the parameter $q$ for hcp Fe.

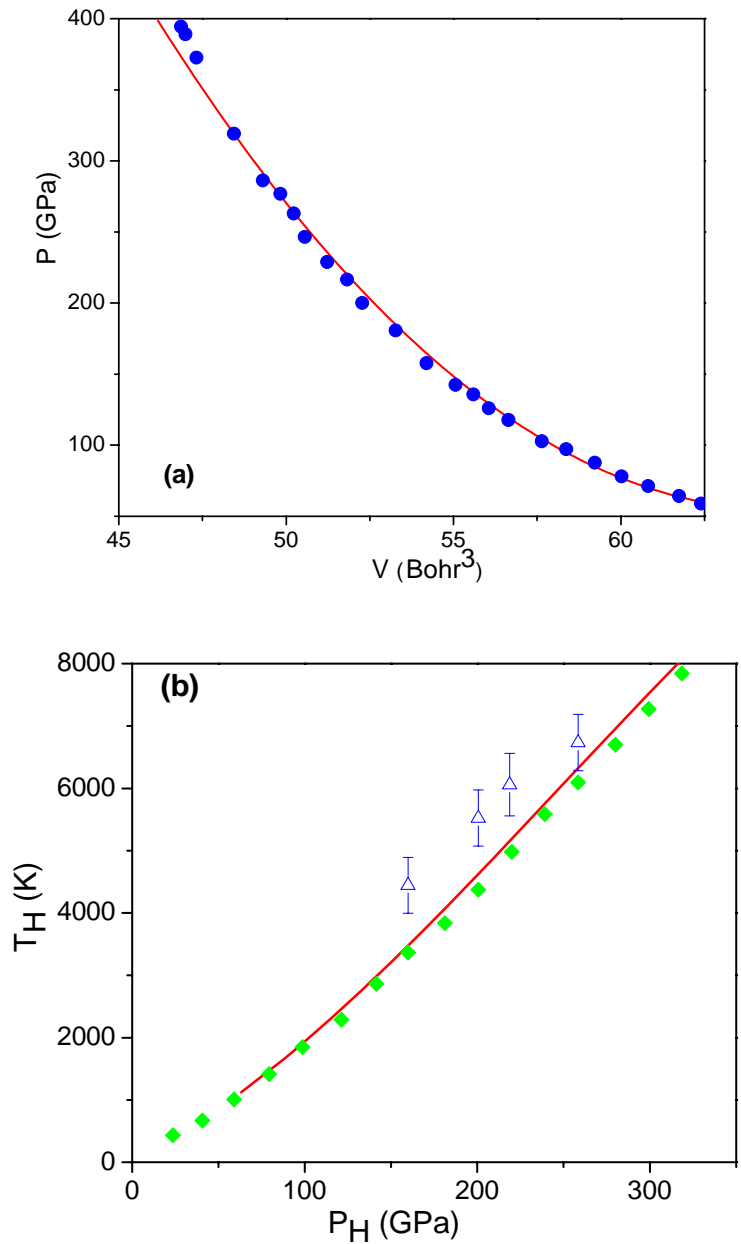

Fig. 10 Shock Hugoniot (a) as well as the temperatures along the Hugoniot (b) for hcp Fe. First-principles data are denoted as lines, in comparison to the shock experiment (filled circles, Ref. 63; open triangles with error bars, Ref. 64) and earlier calculated Hugoniot temperatures (filled diamonds, Ref. 65).

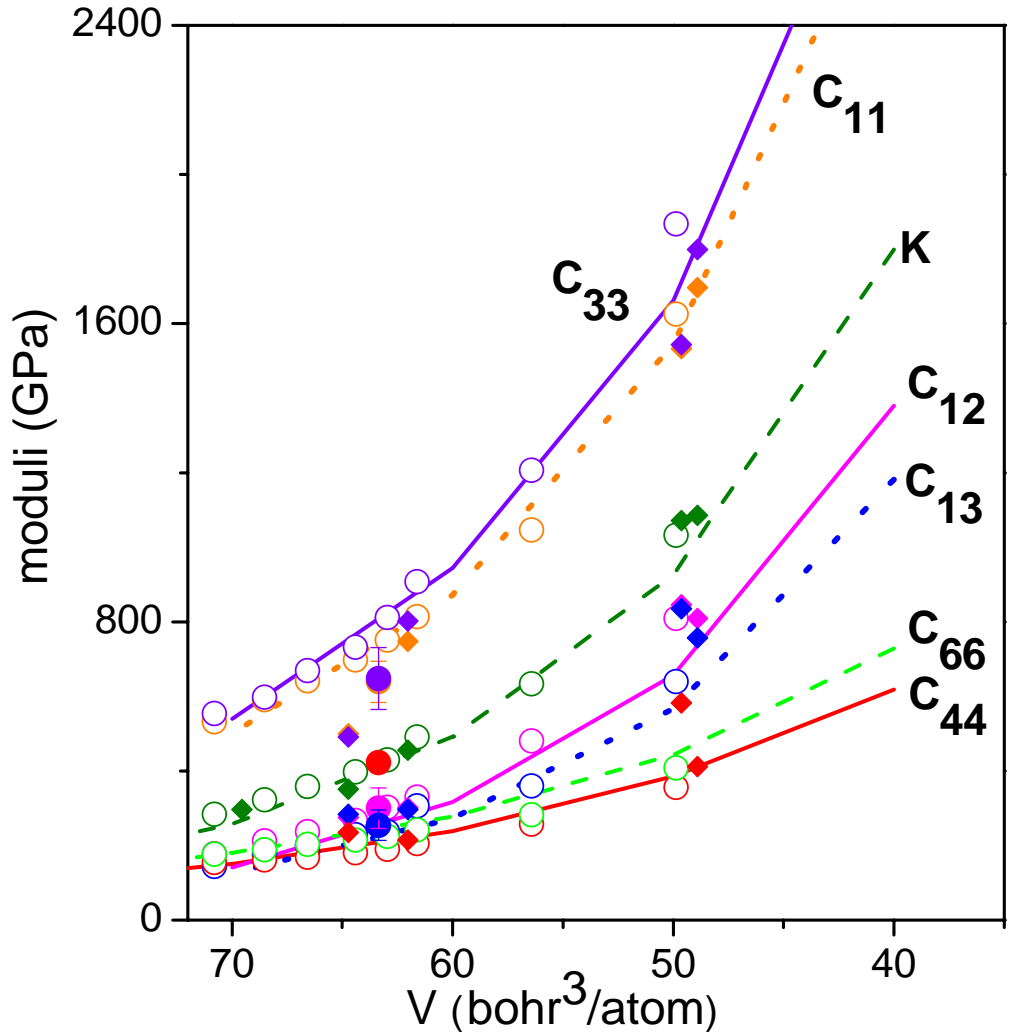

Fig. 11 Static elastic and bulk moduli of hcp Fe (lines) as a function of atomic volume, in comparison to recent theoretical results using augmented-plane-wave plus local orbital method (open circles, Ref. 27), and ambient-temperature experimental data using X-ray diffraction and ultrasonic techniques (filled diamonds, Ref. 71 & 72; filled circles with error bars, Ref. 73).

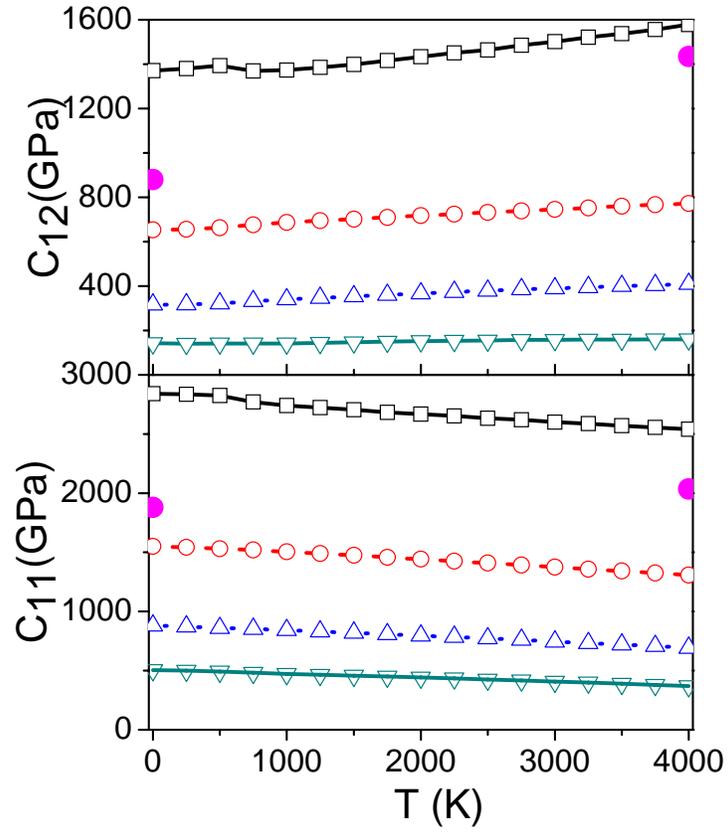
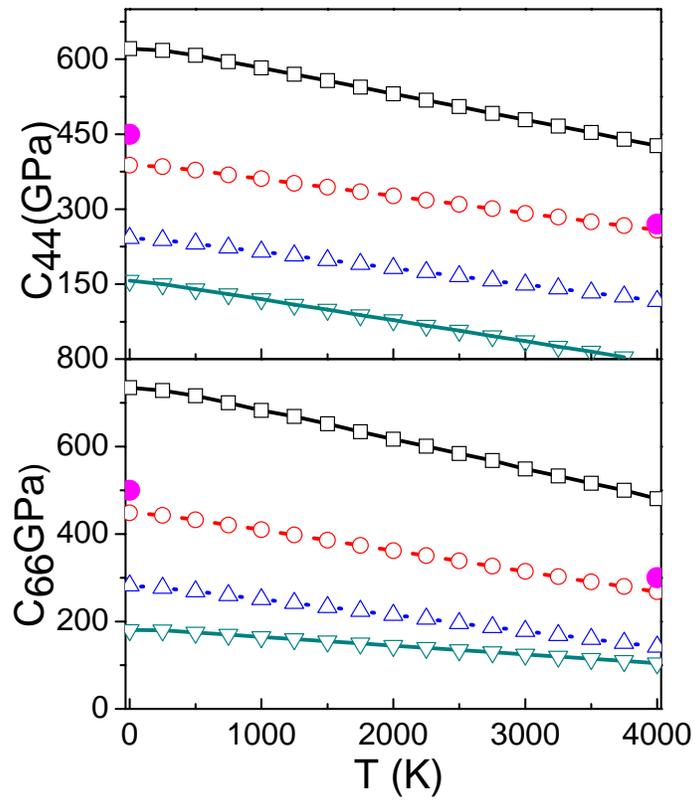

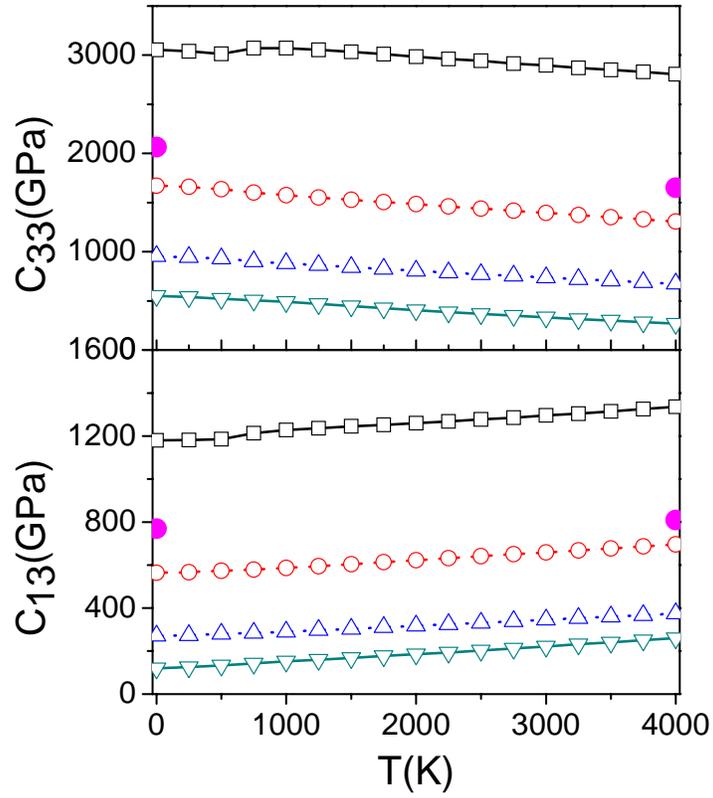

Fig. 12 The calculated temperature dependences of the elastic moduli for nonmagnetic hcp Fe, at atomic volumes from 40 (uppermost curve) to 70 bohr$^3$/atom (lowest curve) in intervals of 10 bohr$^3$/atom. Previous first-principles results using a plane-wave mixed basis method and PIC model at atomic volume of 48 bohr$^3$/atom (filled circles, Ref. 20) are also shown.

Table I  Ambient equation-of-state parameters for hcp Fe determined by fitting a Vinet equation to the free energies. The theoretical calculations are performed on nonmagnetic hcp Fe, except two antiferromagnetic configurations denoted as afmI and afmII.

|  | $V_0$ (bhor$^3$) | $K_0$ (GPa) | $K_0$' |
|---|---|---|---|
| LMTO-GGA | 68.1 | 296 | 4.4 |
| expt (Ref. 52) | 75.4 | 165 | 5.33 |
| expt (Ref. 53) | 75.6 | 156 | 5.81 |
| expt (Ref. 54) | 75.7 | 163.4 | 5.38 |
| LAPW-GGA (Ref. 38) | 69.0 | 292 | 4.4 |
| PAW-GGA (Ref. 51) | 69.2 | 293 |  |
| LMTO-GGA (Ref. 26) | 65.5 | 340 |  |
| LAPW-LDA (Ref. 38) | 64.7 | 344 | 4.4 |
| afmI, LAPW-GGA (Ref. 38) | 70.5 | 210 | 5.5 |
| afmII, LAPW-GGA (Ref. 38) | 71.2 | 209 | 5.2 |